\documentclass[%
reprint,
 amsmath,amssymb,
 aps,
 prd,
] {revtex4-1}

\usepackage{graphicx}% Include figure files
\usepackage{dcolumn}% Align table columns on decimal point
\usepackage{bm}% bold math
\usepackage{color}
\usepackage{ulem}
\begin{document}

%\preprint{APS/123-QED}

\title{The holographic s+p model in 4D and 5D Einstein-Gauss-Bonnet gravity}

\author{Xing-Kun Zhang}
\affiliation{Center for gravitation and astrophysics, Kunming University of Science and Technology, Kunming 650500, China}
\affiliation{College of Physics, Nanjing University of Aeronautics and Astronautics, Nanjing 211106, China}
\author{Zhang-Yu Nie}
 \email{niezy@kust.edu.cn}
 \email{Corresponding author.}
\affiliation{Center for gravitation and astrophysics, Kunming University of Science and Technology, Kunming 650500, China}
\author{Hui Zeng}
 \email{zenghui@kust.edu.cn}
 \email{Corresponding author.}
\affiliation{Center for gravitation and astrophysics, Kunming University of Science and Technology, Kunming 650500, China}
\author{Qiyuan Pan}
 \email{panqiyuan@hunnu.edu.cn}
 \email{Corresponding author.}
\affiliation{Key Laboratory of Low Dimensional Quantum Structures and Quantum Control of Ministry of Education, Synergetic Innovation Center for Quantum Effects and Applications, and Department of Physics, Hunan Normal University, Changsha, Hunan 410081, China}
\date{\today}

\begin{abstract}
We study the holographic s+p model in both four dimensional (4D) and five dimensional (5D) Einstein-Gauss-Bonnet (EGB) gravity. We first show a phase diagram with the Gauss-Bonnet parameter fixed to a small value $\alpha=10^{-7}$ to choose propitiate values of $q_p/q_s$. Then we fix the value of $q_p/q_s$ and plot $\alpha-\mu$ phase diagrams to show the influence of Gauss-Bonnet term on the phase transitions in both 4D and 5D bulk, respectively. The phase diagrams in 4D and 5D present the same qualitative features, indicating similarity of 4D Einstein-Gauss-Bonnet gravity with the 5D case in holography. We also study the influences of Gauss-Bonnet parameter on the special values of the fourth order nonlinear term parameters $\lambda_s$ and $\lambda_p$, below which the condensate grows to a different direction near the critical point, that is important in realizing 1st order superfluid phase transitions. Especially, we notice that these special values are different in the canonical and grand canonical ensembles, which is closely related to the study of the spinodal region, where the phase separations occurs with the linear instability at finite wave vector.

%\begin{description}
%xxx
%xxx
%\end{description}
\end{abstract}
\maketitle

%\tableofcontents

\section{Introduction}
In recent years, the anti-de Sitter/conformal field theory (AdS/CFT) correspondence~\cite{Maldacena:1997re} has attracted a lot attentions and is one of the most remarkable achievements in string theory. Based on this duality, the holographic dual of superconductor phase transitions was set up with a simple gravitational system~\cite{Gubser:2008px,Hartnoll:2008vx}. This novel study was further extended to different order parameters such as the p-wave and d-wave ones by introducing different charged material fields in the gravity side~\cite{Gubser:2008wv,Cai:2013pda,Cai:2013aca,Chen:2010mk,Benini:2010pr}. Holographic models with multi-condensate were also investigated in the last decade~\cite{Cai:2015cya,Basu:2010fa,Cai:2013wma,Nie:2013sda,Nie:2014qma,Nie:2015zia,Amado:2013lia,Arias:2016nww,Nishida:2014lta,Li:2014wca,Nie:2016pjt,Li:2017wbi,Nie:2020lop,Xia:2021pap,Zhang:2021vwp,Yang:2021mit} for better understanding of competition and coexistence of different orders.

Einstein's equations are subjected to curvature square corrections classically in more general theories of gravity. Among the many theories, the Einstein-Gauss-Bonnet theory proposed by Lanczos~\cite{Cornel Lanczos,Cornel Lanczos2} is remarkably simple, for the equations of motion still involve only second order derivatives in the equations of motion.
Within the Einstein-Gauss-Bonnet gravity, the holographic system usually gets nontrivial $\alpha$ corrections, for example, the shear viscosity over entropy ratio breaks the KSS bound~\cite{Kovtun:2004de,Brigante:2007nu,Brigante:2008gz,Cai:2008ph}. Moreover, with larger value of the Gauss-Bonnet parameter $\alpha$, the holographic superconductors usually get lower value of critical temperature and both the scalar~\cite{Gregory:2009fj,Pan:2009xa,Barclay:2010up,Brihaye:2010mr,Gregory:2010yr,Kanno:2011cs} and vector ~\cite{Cai:2010cv,Pan:2011ah,Li:2011xja,Lu:2016smd,Liu:2016utt} hairs are harder to condensate.

It was found that the space dimension should be larger than 4 under the limitations of the Mermin-Wagner theorem. However, Glavan and Lin rescaled the coupling constant and proposed the Einstein-Gauss-Bonnet theory in four dimensional spacetime~\cite{Glavan:2019inb}. Therefore in recent studies, many attentions were paid to the effect of curvature corrections on the holographic superfluid models in the 4D Einstein-Gauss-Bonnet gravity~\cite{Qiao:2020hkx,Ghorai:2021uby,Pan:2021jii}.

However, most works focus on the dependence of the critical point on the Gauss-Bonnet parameter in models with single condensate. It is important to also extend the study of multi-condensate, including competition and coexistence between different orders, to Einstein-Gauss-Bonnet gravity to see possible nontrivial effects. The competition and coexistence between one s-wave and one p-wave order from the 5D Einstein-Gauss-Bonnet gravity were studied in Ref.~\cite{Nie:2015zia}, with a special interest on the pressure-temperature (P-T) phase diagram. Then the holographic system with two s-wave orders from the 5D Einstein-Gauss-Bonnet gravity was studied in Ref.~\cite{Li:2017wbi}, where a reentrant phase transition involving the s+s phase was realized in a systematic way near the Chern-Simons limit. In a recent study~\cite{Qiao:2020hkx}, the authors studied the holographic s-wave and p-wave superconductors respectively in the 4D Einstein-Gauss-Bonnet gravity, showing that the larger value of Gauss-Bonnet parameter always makes the superconductor phase transition harder, which is also a criterion to confirm that the fixed value of operator dimension is a better choice than the fixed mass parameter in studying the higher curvature corrections. However, in this study~\cite{Qiao:2020hkx}, the competition and coexistence between the s-wave and p-wave orders were not considered. And in the previous study of holographic s+p model in 5D~\cite{Nie:2015zia}, the SU(2) p-wave model only contains the $m_p^2=0$ and $q_p=1$ results of a more general p-wave model~\cite{Nie:2014qma}, and the phase structures were only presented in some discrete values of Gauss-Bonnet parameter $\alpha$ in the causality bound. It is necessary to further explore the influence of the higher curvature corrections on the phase transitions involving two competing orders in the holographic s+p model in the 4D Einstein-Gauss-Bonnet gravity and extend the previous 5D results with SU(2) p-wave model to the massive vector p-wave model~\cite{Cai:2013pda,Cai:2013aca} as a good comparison. It is also interesting to see whether similar results of reentrant phase transitions are available as in the s+s model near the Chern-Simons limit~\cite{Li:2017wbi} and explore the possible universality.

In a recent study on the holographic s+p model~\cite{Zhang:2021vwp}, nonlinear terms were included and showed great power on tuning the competition and coexistence of the two orders, which is believed to be universal because of the consistency with physical understanding of these nonlinear terms as (self-)interactions.
%and need further tested in more general holographic systems. 
Later in Ref.~\cite{Zhao:2022jvs}, the fourth and sixth nonlinear terms were considered in the holographic model with single s-wave order, where more general phase transitions including the 0th order and 1st order phase transitions were realized and the linear stability was studied. It was found that there is an important special negative value for the parameter $\lambda$ of the fourth order term, below which the condensate grows to an opposite direction~\cite{Herzog:2010vz} and a first order superfluid phase transition is available with an additional positive value for the parameter of the sixth order term. It is therefore also interesting to further study the influence of higher curvature term on this special value of $\lambda$ to prepare necessary information for realizing various phase transitions.

In this paper, we study the holographic s+p model in the 4D and 5D Einstein-Gauss-Bonnet gravity in the probe limit. We first study the influence of the Gauss-Bonnet parameter on the phase transitions involving competition and coexistence of the two different orders. After that, we also study the influence of Gauss-Bonnet term on the special value of the fourth order potential terms for both the s-wave and the p-wave orders in canonical and grand canonical ensembles, respectively. The rest of this paper is organized as follows. In Sec.~\ref{sect:setup}, we give the setup of the holographic model. In Sec.~\ref{diagram figure}, we show the phase structure both in the 4D and 5D Gauss-Bonnet gravity. In Sec.~\ref{different ensemble}, we study the dependence of the special values of the fourth order potential term parameters on the Gauss-Bonnet parameter for both the s-wave and p-wave orders in the canonical ensemble as well as the grand canonical ensemble. We conclude and discuss our results and some future topics in Sec.~\ref{conclusion}.
\section{Holographic model of an s+p superconductor}  \label{sect:setup}
\subsection{The model setup}
The action for the holographic s+p model is
\begin{eqnarray}
S&=&S_{M}+S_{G}~,\\
S_G&=&\frac{1}{2\kappa_g ^2}\int d^{D}x\sqrt{-g}(R-2\Lambda+\bar{\alpha} \mathcal{R}_{\mathrm{GB}}^{2})~,
\\
S_M&=&\int d^{D}x\sqrt{-g}\Big(-\frac{1}{4}F_{\mu\nu}F^{\mu\nu}
-D_{\mu}\psi^{\ast}D^{\mu}\psi-m^{2}_{s}\psi^{\ast}\psi \nonumber \\ \ &&  \quad
-\frac{1}{2}\rho^{\dagger}_{\mu\nu}\rho^
{\mu\nu}-m^{2}_{p}\rho^{\dagger}_{\mu}\rho^{\mu} 
%\color{blue}{-\lambda_{sp}(\psi^{\ast}\psi)(\rho^{\dagger}_{\mu}\rho^{\mu})}
\nonumber \\ \ &&  \quad
-\lambda_{s}(\psi^{\ast}\psi)^{2}-\lambda_{p}(\rho^{\dagger}_{\mu}\rho^{\mu})^{2}
\Big),
\end{eqnarray}
where $\mathcal{R}_{\mathrm{GB}}^{2}=\mathcal{R}^{2}-4 \mathcal{R}_{\mu \nu} \mathcal{R}^{\mu \nu}+\mathcal{R}_{\mu \nu \rho \sigma} \mathcal{R}^{\mu \nu \rho \sigma}$ is the Gauss-Bonnet term with the coefficient $\alpha$. $F_{\mu\nu}=\nabla_{\mu}A_{\nu}-\nabla_{\nu} A_{\mu}$ is the Maxwell field strength. $\psi$ is the complex scalar field and $\rho_{\mu}$ is complex vector field, with the covariant derivatives $D_{\mu}\psi=\nabla_{\mu}\psi-iq_{s}A_{\mu}\psi$ and $\rho_{\mu\nu}=\bar{D}_{\mu}\rho_{\nu}-\bar{D}_{\nu}\rho_{\mu}$, where $\bar{D}_{\mu}=\nabla_{\mu}-iq_{p}A_{\mu}$. $\Lambda=-\frac{(D-1)(D-2)}{2 L^2}$ is the negative cosmological constant and $L$ is the AdS radius. The two nonlinear terms with coefficients $\lambda_s$ and $\lambda_p$ are useful in tuning the phase structure~\cite{Zhang:2021vwp}, as well as in realizing the 0th and 1st order phase transitions~\cite{Zhao:2022jvs}. 

We work in the probe limit, and take the background spacetime to be the planar symmetric black-brane solution with the line element~\cite{Cai:2001dz}
\begin{eqnarray}\label{metric}
d s^{2}=-f(r) d t^{2}+\frac{d r^{2}}{f(r)}+r^{2} \sum_{i=1}^{D-2} d x_{i}^{2}~,
\end{eqnarray}
where the function $f(r)$ is
\begin{eqnarray}\label{metricf}
f(r)=\frac{r^2}{2\alpha}\bigg[1-\sqrt{1-\frac{4\alpha}{L^2}\Big(1-\frac{r_h^{(D-1)}}{r^{(D-1)}}\Big)}~\bigg]~.
\end{eqnarray}
The new Gauss-Bonnet coefficient $\alpha$ is related to $\bar{\alpha}$ in the action with the relation $\alpha=\bar{\alpha} (D-3)(D-4)$~\cite{Cai:2001dz}. For the case of $D=4$, this means the original parameter $\bar{\alpha}$ should be infinitely large in order to get a finite contribution of the Gauss-Bonnet term and a finite value of $\alpha$. Nevertheless, the solution is consistent with a direct $D\rightarrow 4$ version of Eq.~(\ref{metric}) with $f(r)$ given by Eq.~(\ref{metricf})~\cite{Glavan:2019inb}.

Near the AdS boundary, the function $f(r)$ get the asymptotic value
\begin{eqnarray}
f(\infty)=\frac{r^2}{2\alpha}\bigg(1-\sqrt{1-\frac{4\alpha}{L^2}}~\bigg)~,
\end{eqnarray}
from which we see that the AdS radius is deformed to the effective value~\cite{Cai:2001dz}
\begin{eqnarray}\label{lengthscale}
L_{\text {eff }}^{2}=\frac{2 \alpha}{1-\sqrt{1-\frac{4 \alpha}{L^{2}}}} \rightarrow\left\{\begin{array}{ll}
L^{2}, & \text { for } \alpha \rightarrow 0~; \\
\frac{L^{2}}{2}, & \text { for } \alpha \rightarrow \frac{L^{2}}{4}~.
\end{array}\right.
\end{eqnarray}
To keep real value of $L_{\text{eff}}^2$, $\alpha\leq L^2/4$ is required, where the upper bound %$\alpha\rightarrow L^2/4$ 
is called the Chern-Simons limit.

The Hawking temperature of this black-brane solution is given by
\begin{eqnarray}
T=\frac{f^{\prime}\left(r_{h}\right)}{4 \pi}=\frac{(D-1) r_{h}}{4 \pi}~,
\end{eqnarray}
%看看1505,12289
with $r_{h}$ the horizon radius.

It is consistent to take the following ansatz~\cite{Nie:2014qma,Nie:2016pjt,Nie:2020lop,Xia:2021pap,Zhang:2021vwp}
\begin{eqnarray}
\psi=\psi_{s}(r),~ A_{t}=\phi(r),~ \rho_{x}=\psi_p(r),
\end{eqnarray}
with all other matter field components set to zero to study the competition and coexistence between the s-wave and p-wave orders. Consequently, the equations of motion of matter fields are
\begin{eqnarray}
\phi''+\frac{D-2}{r}\phi'-2\Big(\frac{q^{2}_{s}\psi^{2}_{s}}{f}+\frac{q^{2}_{p}\psi^{2}_{p}}{r^{2}f}\Big)\phi=0,\label{EqPhi}\\
\psi_{p}''+\Big(\frac{f'}{f}+\frac{D-4}{r}\Big)\psi_{p}'+\Big(\frac{q_{p}^{2}\phi^{2}}{f^{2}}-\frac{m^{2}_{p}}{f}\Big)\psi_{p}-\frac{2\lambda_{p}}{r^2 f}\psi^{3}_{p}=0,\label{EqPsip}\\
\psi_{s}''+\Big(\frac{f'}{f}+\frac{D-2}{r}\Big)\psi_{s}'+\Big(\frac{q_{s}^{2}\phi^{2}}{f^{2}}-\frac{m^{2}_{s}}{f}\Big)\psi_{s}-\frac{2\lambda_{s}}{f}\psi^{3}_{s}=0.\label{EqPsis}
\end{eqnarray}
The Maxwell field should get finite norm at the horizon, which implies $\phi(r=r_h)$=0. Therefore the expansions of the three functions near the horizon are \begin{eqnarray}
&\phi(r)=\phi_{1}(r-r_{h})+\mathcal{O}((r-r_{h})^{2}),\\
&\psi_{s}(r)=\psi_{s0}+\psi_{s1}(r-r_{h})+\mathcal{O}((r-r_{h})^{2}),\\
&\psi_{p}(r)=\psi_{p0}+\psi_{p1}(r-r_{h})+\mathcal{O}((r-r_{h})^{2}).
\end{eqnarray}
The charged scalar field and the charged vector field get natural boundary conditions at the horizon, which further implies $\psi_{s0}=(D-1)\psi_{s1}/m_s^2$ and $\psi_{p0}=(D-1)\psi_{p1}/m_p^2$ for finite solutions, respectively.

The expansions of the three fields near the AdS boudnary are
\begin{eqnarray}
&\phi(r)=\mu-\frac{\rho}{r^{D-3}}+...~,\\
&\psi_{p}=\frac{\psi_{p_{-}}}{r^{\Delta_{p-}}}+\frac{\psi_{p_{+}}}{r^{\Delta_{p+}}}+...~,\\
&\psi_{s}=\frac{\psi_{s_{-}}}{r^{\Delta_{s-}}}+\frac{\psi_{s_{+}}}{r^{\Delta_{s+}}}+...~,
\end{eqnarray}
where
\begin{eqnarray}\label{boundary dimension s}
\Delta_{s \pm}=\frac{1}{2}\left[(D-1) \pm \sqrt{(D-1)^{2}+4 m_s^{2} L_{\text {\text{eff}}}^{2}}\right],\\
\Delta_{p \pm}=\frac{1}{2}\left[(D-3) \pm \sqrt{(D-3)^{2}+4 m_p^{2} L_{\text {\text{eff}}}^{2}}\right].
\end{eqnarray}
%and $m^{2} L_{\text {eff }}^{2}$ call the effect mass.
$\mu$ and $\rho$ are the chemical potential and charge density of the boundary field theory, respectively. $\Delta_{s \pm}$ and $\Delta_{p \pm}$ are the conformal dimensions for the sources and expectation values for the s-wave and p-wave orders. $m_s^2 \geq -{(D-1)^2}/{(4 L_{\text{eff}}^2)}$ and $m_p^2 \geq -{(D-3)^{2}}/{(4 L_{\text{eff}}^2}$) are known as the Breitenlohner-Freedman bound~\cite{Breitenlohner:1982bm}.

We choose the standard quantization, which means that $\psi_{s-}$ and $\psi_{p-}$ are the sources of the s-wave and p-wave orders while $\psi_{s+}$ and $\psi_{p+}$ are the expectation values. In order to obtain the superfluid solutions with the spontaneous symmetry breaking, we impose the source free boundary conditions $\psi_{s-}=\psi_{p-}=0$. Then with a controlled value of the chemical potential $\mu$, we fix all the degrees of freedom for the group of three coupled second order differential equations.

There are three scaling symmetries in these coupled equations. The first set of the scaling symmetry~(\ref{scaling1}) is used to rescale the matter fields as well as the charge parameters $q_s$ and $q_p$. As a result, the value of $q_s$ can always be rescaled to 1, and only the ratio $q_p/q_s$ changes the phase structure qualitatively. Therefore we set $q_s=1$ and further use the second and third sets of scaling symmetries~(\ref{scaling2}) and (\ref{scaling3}) to set $L=r_h=1$ in the rest of this work without loss of generality. 

\begin{eqnarray}
(1)& \phi \rightarrow \lambda \phi, \psi_s \rightarrow \lambda \psi_s, \psi_p \rightarrow \lambda \psi_p \nonumber \\
&q_{s} \rightarrow \lambda^{-1} q_{s}, q_{p} \rightarrow \lambda^{-1} q_{p}, \nonumber \\
& \lambda_{s} \rightarrow \lambda^{-2} \lambda_{s}, \lambda_{p} \rightarrow \lambda^{-2} \lambda_{p};\label{scaling1}\\
(2)&\phi \rightarrow \lambda^2 \phi, \psi_s \rightarrow \lambda \psi_s, \psi_p \rightarrow \lambda \psi_p,\nonumber \\
& f \rightarrow \lambda^{2} f, m_{s} \rightarrow \lambda m_{s}, m_{p} \rightarrow \lambda m_{p}, \nonumber \\
&L \rightarrow \lambda^{-1} L, \alpha \rightarrow \lambda^{-2} \alpha; \label{scaling2}\\
(3)& \phi \rightarrow \lambda \phi, \psi_p \rightarrow \lambda \psi_p,\nonumber \\
&f\rightarrow \lambda^{2} f, r\rightarrow \lambda r. \label{scaling3}
\end{eqnarray}
\subsection{Grand potential}
In order to investigate the competition between the different solutions, it is necessary to compare the thermodynamic potential of these solutions. We work in the grand canonical potential in this work and calculate the grand potential from the on-shell Euclidean action. In the probe limit, only the contribution of matter part
\begin{eqnarray}
\Omega_{m}=V_{D-2}  \Big[   -\frac{(D-3)\mu \rho}{2} \nonumber\\
+\int_{r_{h}}^{\infty} (   \frac{q_{s}^{2} r^{D-2} \phi^{2} \psi_{s}^{2}}{f}+\frac{q_{p}^{2} r^{D-4} \phi^{2} \psi_{p}^{2}}{f} \nonumber \\
-\lambda_{s} r^{D-2} \psi_{s}^{4}-\frac{\lambda_{p} r^{D} \psi_{p}^{4} }{r^{6}})dr\Big],
\end{eqnarray}
differs for the different solutions. The prefactor $V_{D-2}$ is the volume of the boundary space.
\section{Phase structure from the 4D and 5D EGB gravity}\label{diagram figure}
In order to present the effects of Gauss-Bonnet term on the phase structure, besides the choice $q_s=L=r_h=1$, it is wise to also fix the other parameters such as $m_s^2$, $m_p^2$ and $q_p$. $m_s^2=-2$ and $m_p^2=0$ are good choices in the 4D Einstein gravity~\cite{Zeng:2019yhi,Xia:2019eje,Li:2020ayr,Zeng:2022hut} while in the 5D case $m_s^2=-3$ and $m_p^2=0$ are better, because the conformal dimensions are integers and full equations in dynamical case are simple. We choose to fix the conformal dimensions rather than mass parameters in this study, because the fixed integer values of conformal dimensions are helpful to simplify the equations, and it is also argued in Ref.~\cite{Qiao:2020hkx} that this is a better choice and gives consistent results that the larger value of Gauss-Bonnet parameter (not close to the Chern-Simons limit) makes the superfluid phase transitions harder. Therefore we choose to fix $m_s^2 L_{\text{eff}}^2=-2$ and $m_p^2 L_{\text{eff}}^2=0$ for different values of Gauss-Bonnet parameter $\alpha$ in the 4D case, while in the 5D cousin, we set $m_s^2 L_{\text{eff}}^2=-3$ and $m_p^2 L_{\text{eff}}^2=0$. For the rest parameter $q_p$, we draw $q_p-\mu$ phase diagrams at $\alpha=1.0\times 10^{-7}$ to help choose better values, which makes the coexistent phase more clear. Both the two nonlinear parameters $\lambda_s$ and $\lambda_p$ are set to zero in this section.

\subsection{$q_p-\mu$ phase diagram at $\alpha=10^{-7}$}
\begin{figure}
\centering
\includegraphics[width=0.49\columnwidth,origin=c,trim=90 260 120 230,clip]{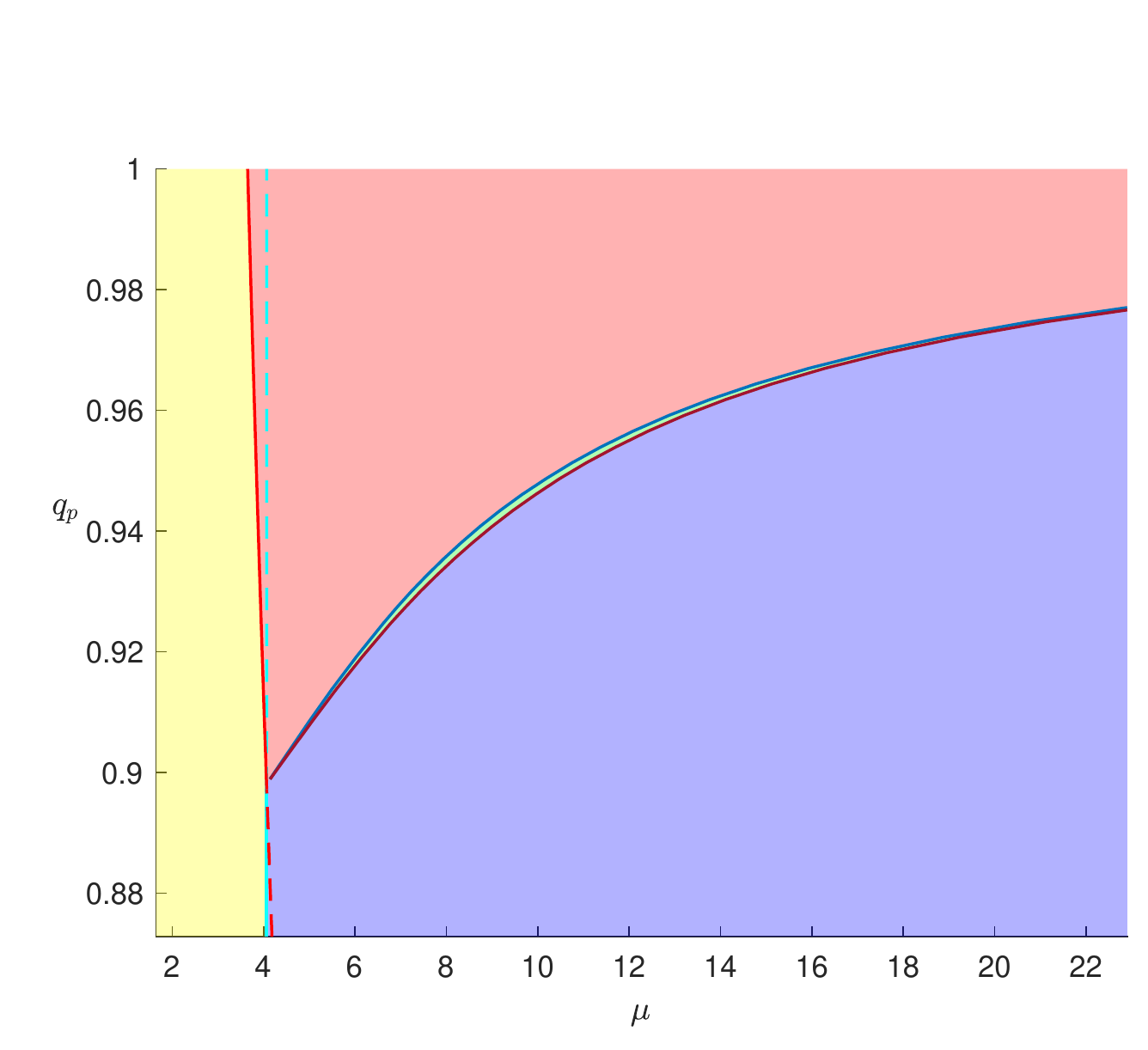}
\includegraphics[width=0.49\columnwidth,origin=c,trim=90 260 120 230,clip]{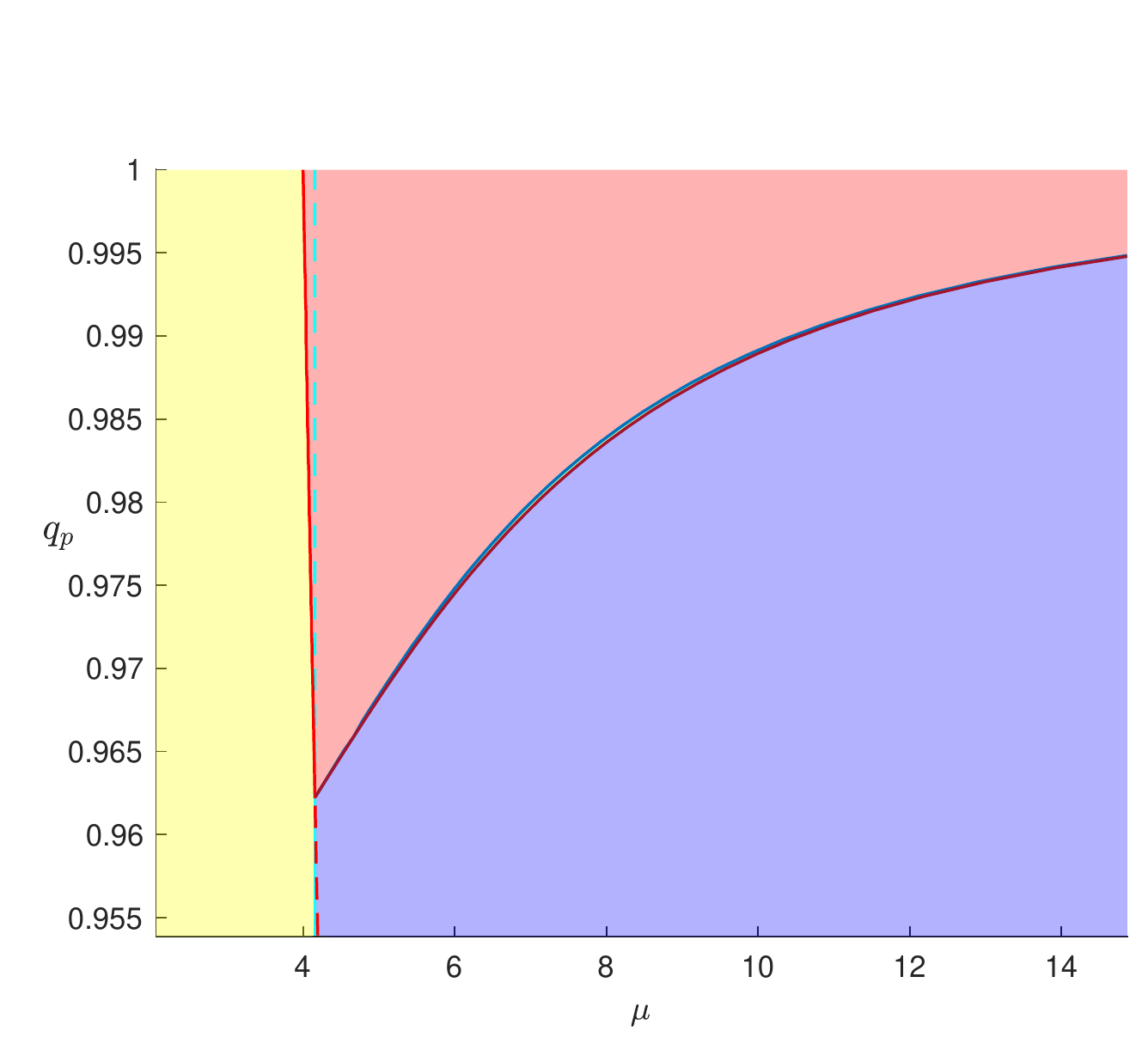}
\caption{The $q_p-\mu$ phase diagrams for the 4D (left) and 5D (right) Einstein-Gauss-Bonnet gravity.
In both the two panels, the following parameters are the same: $\alpha=10^{-7}$, $\lambda_s=\lambda_p=0$, $m_p^2=0$. In the holographic model with 4D bulk we set $m_s^2=-2/L_{\text{eff}}^2$, while in the holographic model with 5D bulk we have $m_s^2=-3/L_{\text{eff}}^2$.
The yellow, red, green and blue regions denote the normal phase, the p-wave phase, the s+p coexistent phase and the s-wave phase respectively.
The light red (blue) line between the yellow and red (blue) region denotes the critical points of the p-wave (s-wave) phase with single condensate. The dark red and dark blue lines on the boundary of the narrow green region denote critical points of the the p-wave and s-wave orders of the coexistence phase, respectively.
}\label{qpmuPD}
\end{figure}

We fix $\alpha=10^{-7}$, which gives the results close to those in the Einstein gravity, and draw the $q_p-\mu$ phase diagram in the 4D and 5D bulk systems in the left and right panels of Fig.~\ref{qpmuPD}, respectively.
%给出图1

From the two panels of Fig.~\ref{qpmuPD}, we can see that both in the holographic models with 4D and 5D bulk, we get qualitatively the same $q_p-\mu$ phase diagrams from the Einstein gravity as well as from the Einstein-Gauss-Bonnet gravity with a very small value of Gauss-Bonnet coefficient. In the two panels, we use yellow, red, blue and green to denote the normal phase, the p-wave phase, the s-wave phase and the s+p coexistent phase, respectively. The coexistent phase is very narrow between the red and blue regions. The light red and light blue lines denote the critical points for the p-wave and s-wave phases, respectively, while the dark red and dark blue lines denote the critical points of p-wave and s-wave orders for the coexist phase, respectively. We can see that in both the 4D and 5D cases, the p-wave phase is more stable in the region with larger value of $q_p$ and smaller value of $\mu$, while the s-wave phase is more stable in the region with smaller $q_p$ and larger $\mu$. Between the two phases, the s+p coexistent phase dominates a narrow region between the two phases with single condensate.

In order to show the coexistent s+p phase more clear in $\alpha-\mu$ phase diagrams in the next subsection, we fix the values of $q_p$ to two special values as $q_{p}=0.9487$ in the model with 4D bulk and $q_{p}=0.9798$ in the model with 5D bulk, respectively.
\subsection{$\alpha-\mu$ phase diagrams}
\begin{figure}
\centering
\includegraphics[width=0.49\columnwidth,origin=c,trim=90 260 120 230,clip]{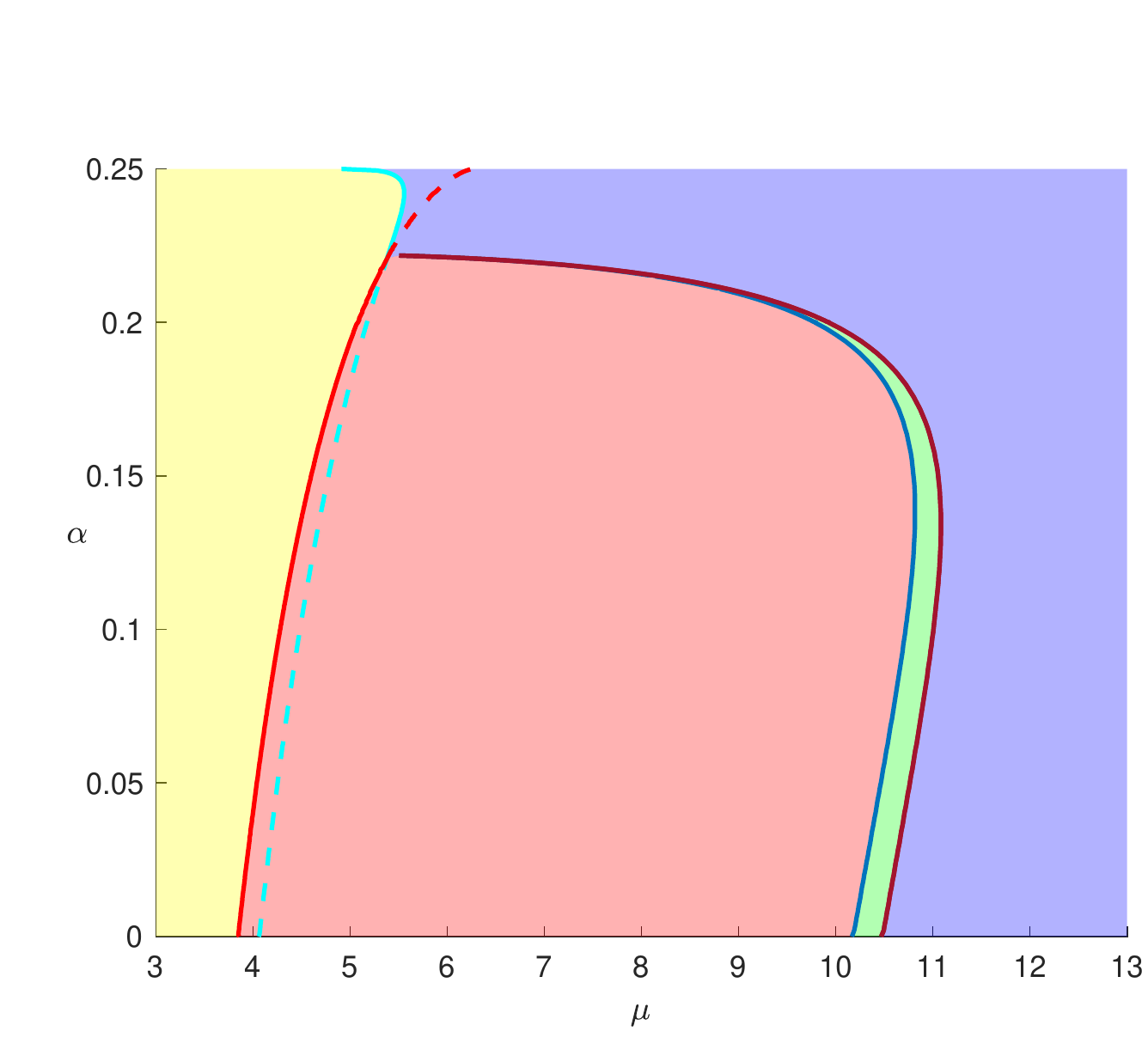}
\includegraphics[width=0.49\columnwidth,origin=c,trim=90 260 120 230,clip]{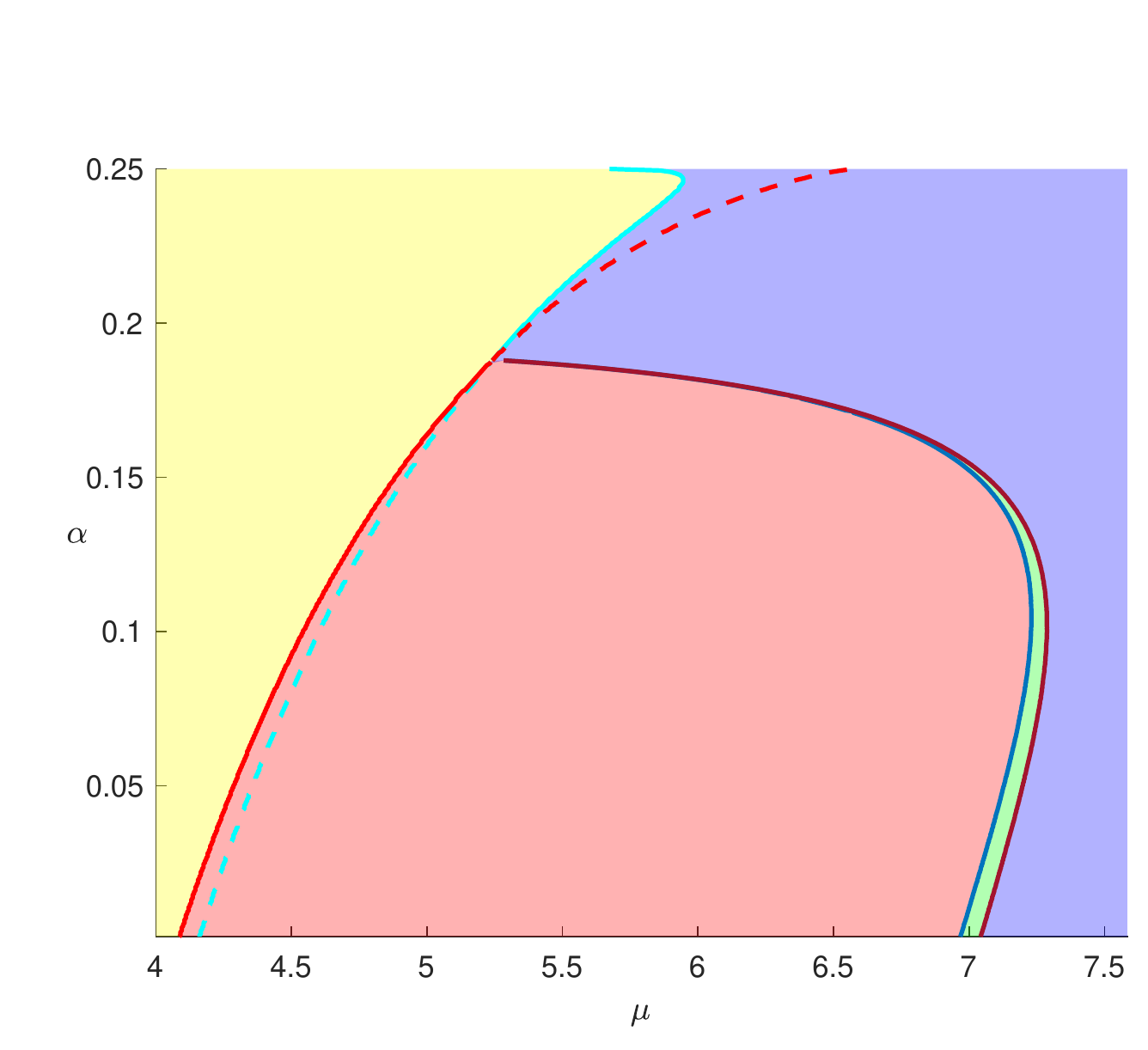}
\caption{The $\alpha-\mu$ phase diagrams of the s+p model with 4D (Left) and 5D (Right) bulk. In the 4D case we set $m_s^2=-2/L_{\text{eff}}^2$, $q_p=0.9487$, and in the 5D case we have $m_s^2=-3/L_{\text{eff}}^2$, $q_p=0.9798$. The notions in these plots are the same as those in Fig.~\ref{qpmuPD}.
}\label{alphamuPD}
\end{figure}

In this section, we investigate the influence of the higher curvature term with parameter $\alpha$ on the phase structure of this s+p model in 4D and 5D bulk spacetime. With $q_p$ fixed to $q_{p}=0.9487$ in the 4D case and $q_{p}=0.9798$ in the 5D case, we draw the $\alpha-\mu$ phase diagrams in Fig.~\ref{alphamuPD}.
All other parameters remain the same as in the previous subsection, and all the notions are the same as those in Fig.~\ref{qpmuPD}. %$\alpha$ near the Chern-Simons limit. 

The left panel of Fig.~\ref{alphamuPD} is the $\alpha-\mu$ phase diagram with $m_s^2=-2/L_{\text{eff}}^2$ and $q_p=0.9487$ from the 4D Einstein-Gauss-Bonnet gravity. We can see from this plot that generally, the critical values of chemical potential for the single condensate s-wave and p-wave phases both increase along the increase of $\alpha$. However, this value for the s-wave phase gets a sudden decreasing near the Chern-Simons limit, which is already observed in Refs.~\cite{Li:2017wbi,Qiao:2020hkx}. It is argued in Ref.~\cite{Li:2017wbi} that this special behavior near the Chern-Simons limit is responsible for the reentrant behavior of the coexistent phase. However, in this phase diagram, all the four critical lines are single valued functions on $\alpha$, and no reentrant behavior is observed at any fixed value of $\alpha$ here. However, we still get non-monotonic dependence on $\alpha$ for the two dark lines, denoting the critical points of the coexistent phase. Along with the increase of $\alpha$, the region dominated by the coexistent phase first moves rightwards with increasing value of $\mu$, then moves leftwards until reaching the intersection point of the two critical points of single condensate solutions. This non-monotonic behavior is closely related to the sudden decreasing of the critical point of s-wave solution near the Chern-Simons limit, which indicates the stability of the s-wave solution increases rapidly near the Chern-Simons limit.

The right panel of Fig.~\ref{alphamuPD} shows the $\alpha-\mu$ phase diagram with $m_s^2=-3/L_{\text{eff}}^2$ and $q_{p}=0.9798$ from the 5D Einstein-Gauss-Bonnet gravity, which is qualitatively the same as the 4D phase diagram in left panel.

The reentrant phase transition of the s+s coexistent phase was realized with $\alpha$ very close to the Chern-Simons limit in Ref.~\cite{Li:2017wbi}. A natural consideration is that if we promote the region dominated by the s+p phase more close to the Chern-Simons limit, it is very likely to realize the reentrant phase transition in the s+p model. Based on the previous $q_p-\mu$ and $\alpha-\mu$ phase diagrams, we are inspired to increase the value of $q_p$ a little bit, which will raise the intersection point of the critical points for two single condensate solutions in the $\alpha-\mu$ phase diagram to a larger value of $\alpha$.

\begin{figure}
\centering
\includegraphics[width=0.49\columnwidth,origin=c,trim=90 260 120 230,clip]{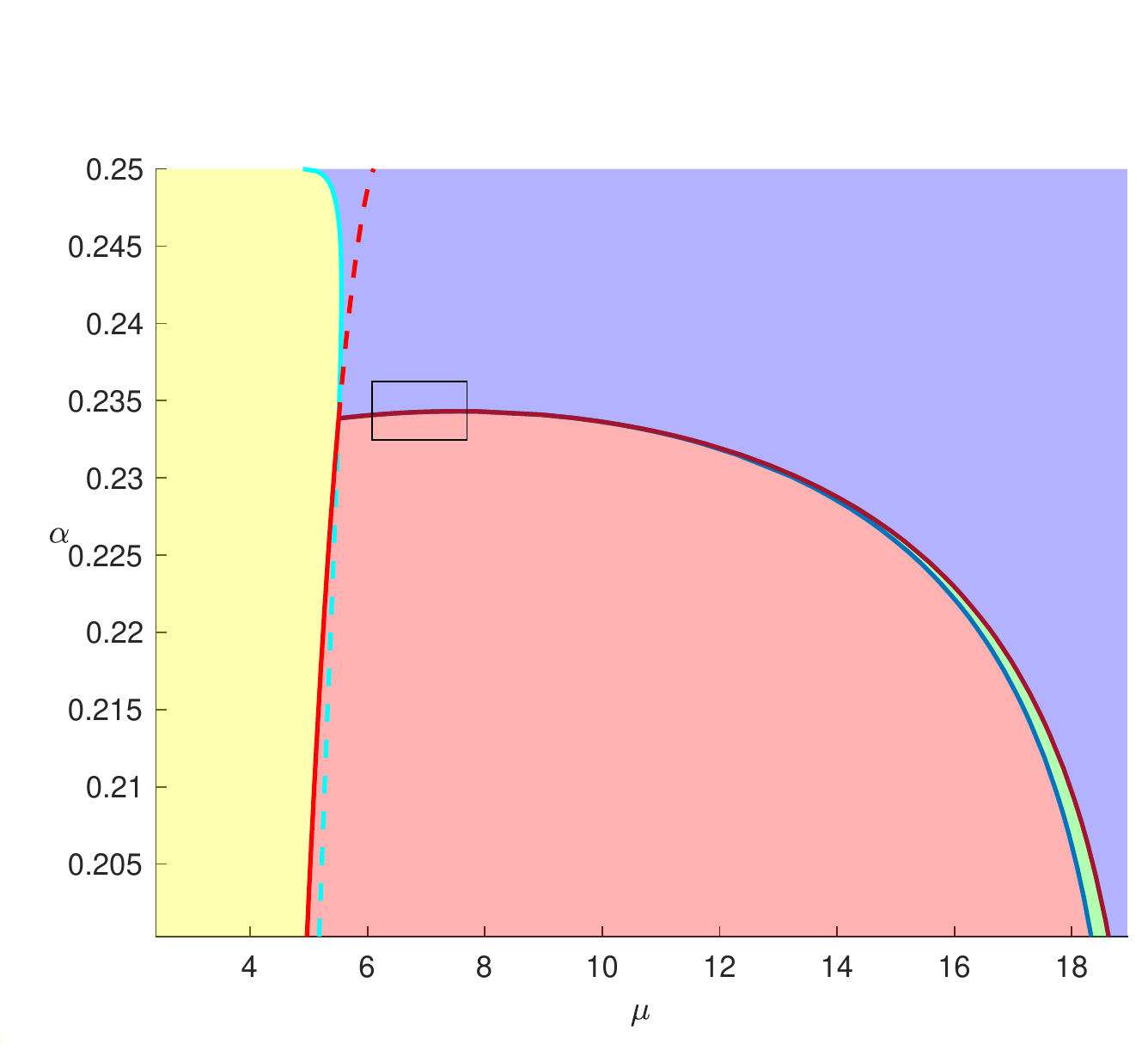}
\includegraphics[width=0.49\columnwidth,origin=c,trim=90 260 120 230,clip]{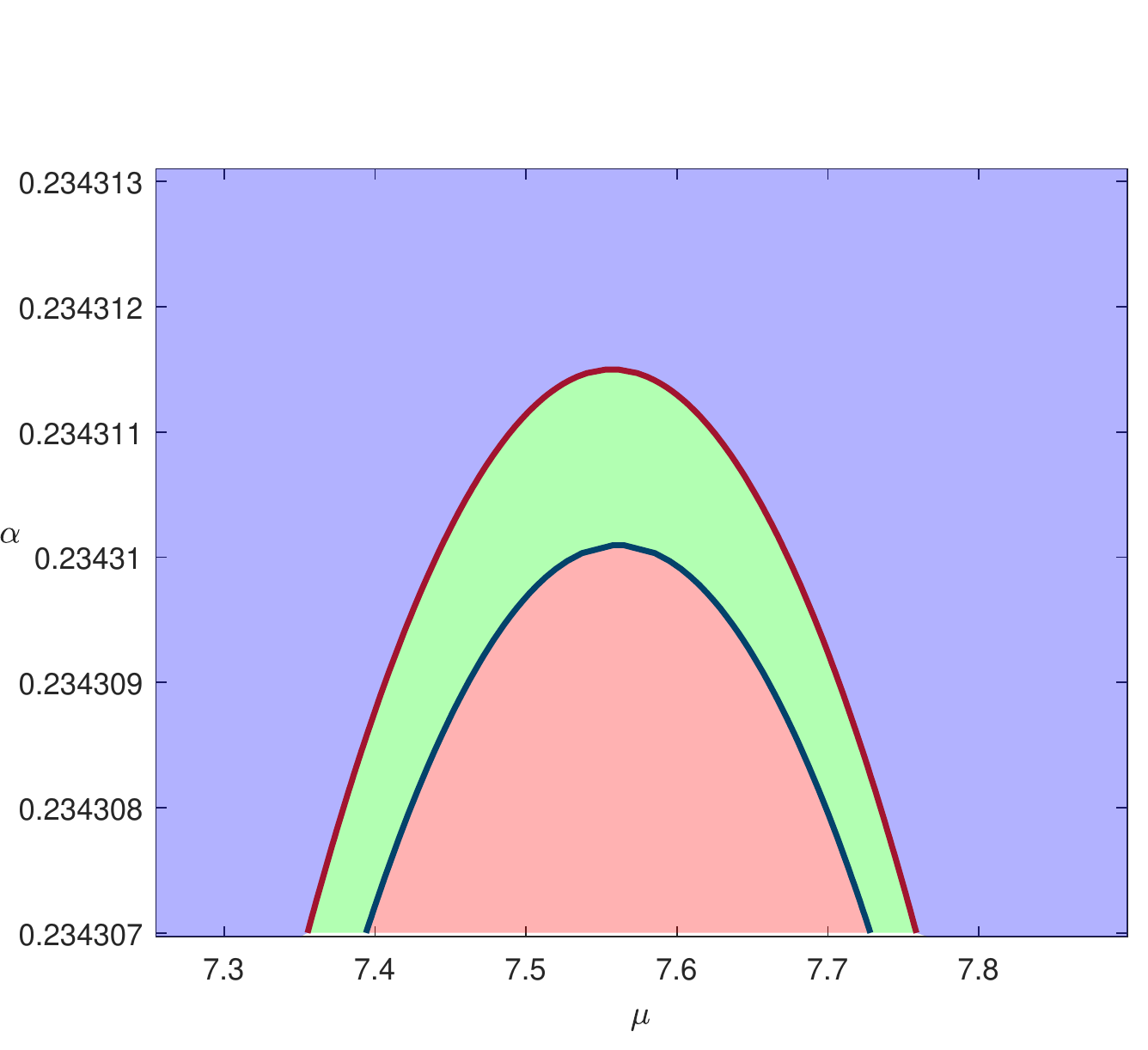}
\caption{
The $\alpha-\mu$ phase diagram with $m_s^2=-2/L_{\text{eff}}^2$, $q_p=0.9695$ from the 4D Einstein-Gauss-Bonnet gravity near the Chern-Simons limit. The right panel show an enlarged version of the black rectangle region in the left panel.
The notions in these plots are the same as those in Fig.~\ref{qpmuPD}.}\label{alphamuPD4DCS}
\end{figure}

We set a new value $q_{p}=0.9695$ and draw the $\alpha-\mu$ phase diagram in the case with 4D bulk in Fig.~\ref{alphamuPD4DCS}, where we focus on the region near the Chern-Simons limit and the right panel is an enlarged version of the rectangle region in the left panel. We can see that when the intersection point of the two critical points for the single condensate solutions is raised to a position near the Chern-Simons limit, the region dominated by the s+p coexistent phase is also raised to this larger value of $\alpha$. Moreover, from the enlarged version we can see that the dark red and dark blue lines show maximum peaks, and a renentrant phase transition of the s+p phase is available with a fixed value of $\alpha$, where the dark red line gets two values of $\mu$~\cite{Li:2017wbi,Xia:2021pap,Zhang:2021vwp}.

\begin{figure}
\centering
\includegraphics[width=0.49\columnwidth,origin=c,trim=90 260 120 230,clip]{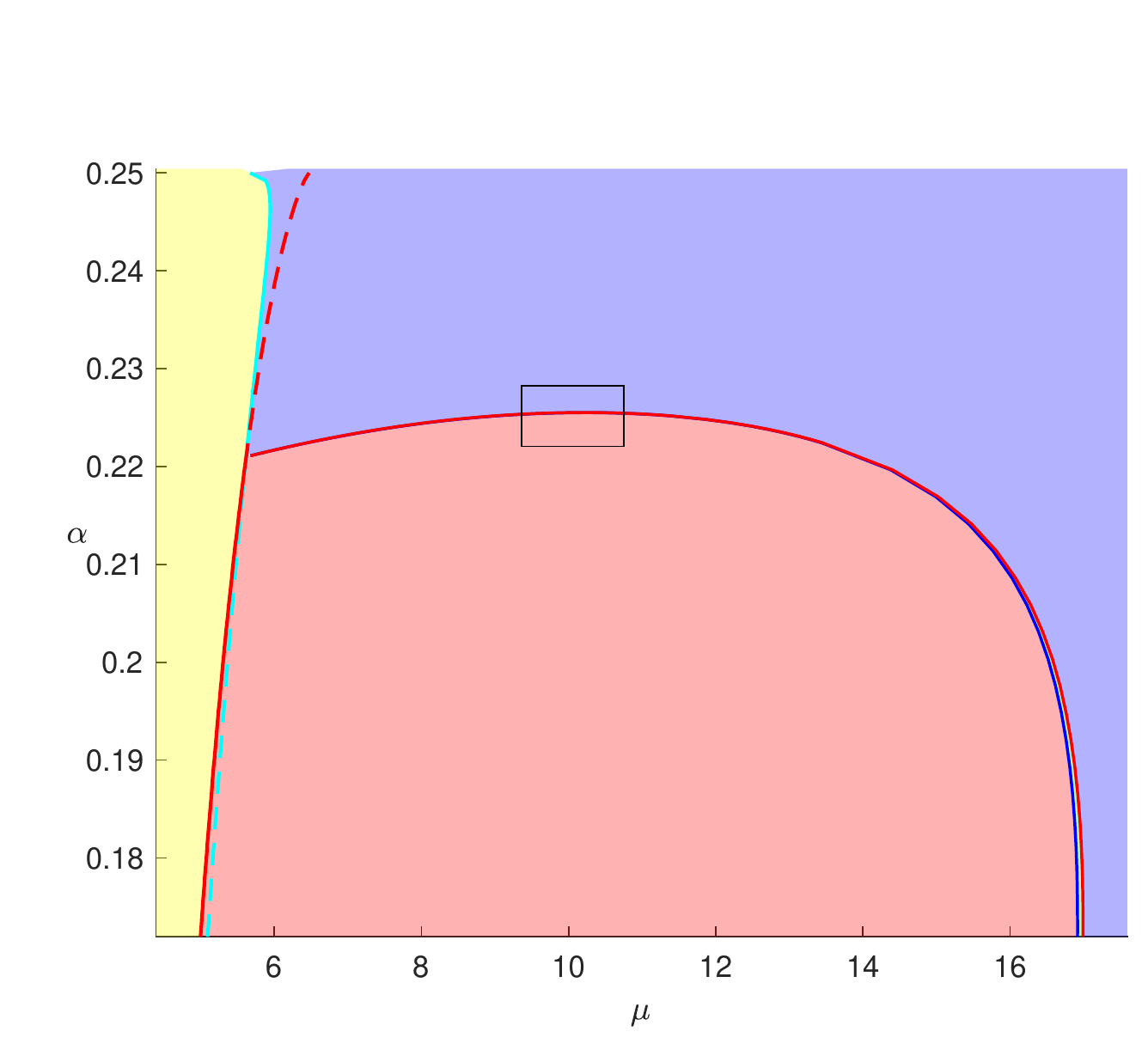}
\includegraphics[width=0.49\columnwidth,origin=c,trim=90 260 120 230,clip]{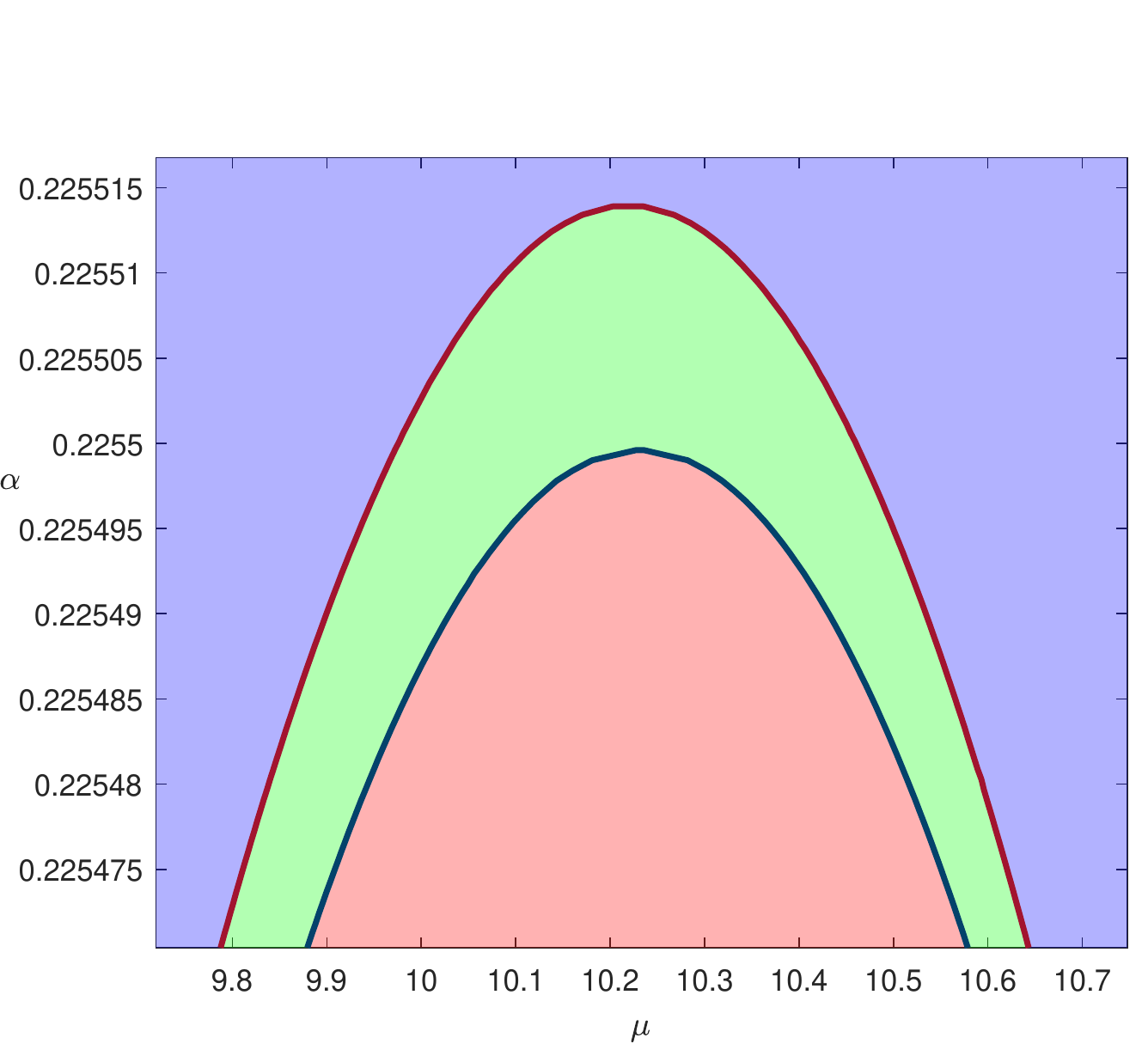}
\caption{
The $\alpha-\mu$ phase diagram with $m_s^2=-3/L_{\text{eff}}^2$, $q_p=0.9950$ from the 5D Einstein-Gauss-Bonnet gravity near the Chern-Simons limit. The right panel show an enlarged version of the black rectangle region in the left panel.
The notions in these plots are the same as those in Fig.~\ref{qpmuPD}.}\label{alphamuPD5DCS}
\end{figure}

We also set a new value of $q_{p}=0.9950$ and plot the $\alpha-\mu$ phase diagram in the case with 5D bulk in Fig.~\ref{alphamuPD5DCS} as a comparison, where we again see qualitatively the same results as in the case with a 4D bulk.
\section{Special values of $\lambda_s$ and $\lambda_p$ in different ensembles and their dependence on the Gauss-Bonnet coefficient $\alpha$}\label{different ensemble}
%为什么要得到0th相变
The influence of the Gauss-Bonnet term on the phase transitions of the holographic s+p system is presented well in the above section. We go on study the dependence of another interesting feature on the Gauss-Bonnet parameter in the holographic s+p model with nonlinear self-interaction terms in this section.

The effect of nonlinear interaction and self-interaction terms is considered in the holographic s+p model in Ref.~\cite{Zhang:2021vwp}, and the nonlinear self-interaction terms are further investigated in the model with single s-wave order in Ref.~\cite{Zhao:2022jvs}. It is very interesting that with a small negative value for the coefficient $\lambda_s$ of the fourth order self-interaction term, the condensate of the single s-wave order shows a typical 0th order phase transition. When the negative value of $\lambda_s$ is lower enough, the condensate grows to an opposite direction at the critical point, and with an additional positive coefficient $\tau_s$ for the sixth order self-interaction term, the s-wave superfluid phase transition becomes 1st order~\cite{Zhang:2021vwp}. Therefore we see a special value for $\lambda_s$ below which the condensate grows leftwards, and this special value is important in realizing the first order superfluid phase transitions.

Furthermore, the authors in Ref.~\cite{Zhao:2022jvs} discovered that the turning point of the condensate curve is different in the canonical ensemble and in the grand canonical ensemble, indicating a section with negative value of susceptibility, which is related to the Cahn-Hilliard instability triggering the spinodal decomposition. Then besides the special value of $\lambda_s$ in the canonical ensemble, it is also necessary to study another special value for $\lambda_s$ in the grand canonical ensemble which is not mentioned in Ref.~\cite{Zhao:2022jvs}. According to the conclusions that the instability section in grand canonical ensemble is larger~\cite{Zhao:2022jvs}, this special value $\lambda_s$ in the grand canonical ensemble should not be lower than the special value in the canonical ensemble.

\begin{figure}
\centering
\includegraphics[width=0.49\columnwidth,origin=c,trim=90 260 120 230,clip]{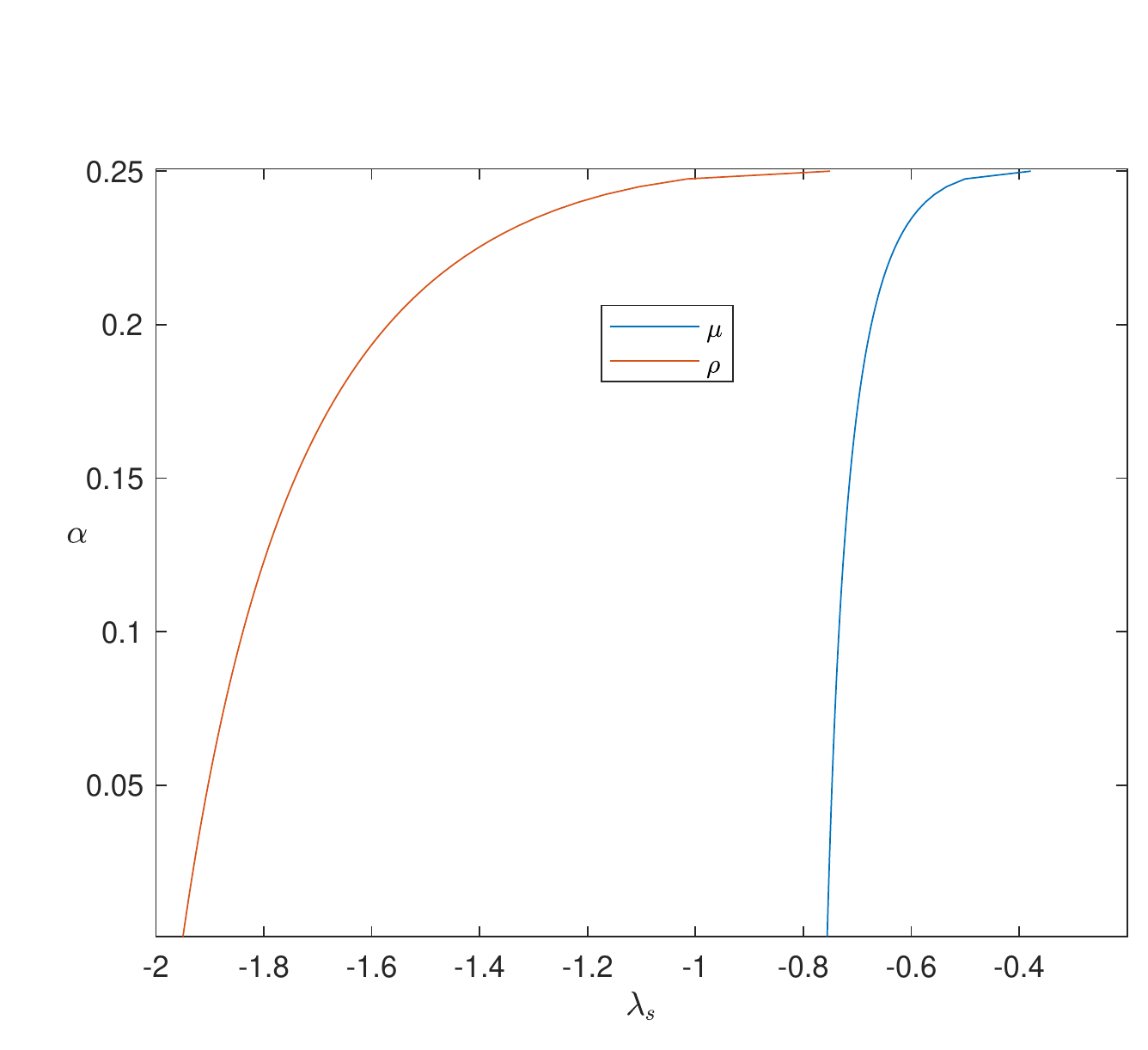}
\includegraphics[width=0.49\columnwidth,origin=c,trim=90 260 120 230,clip]{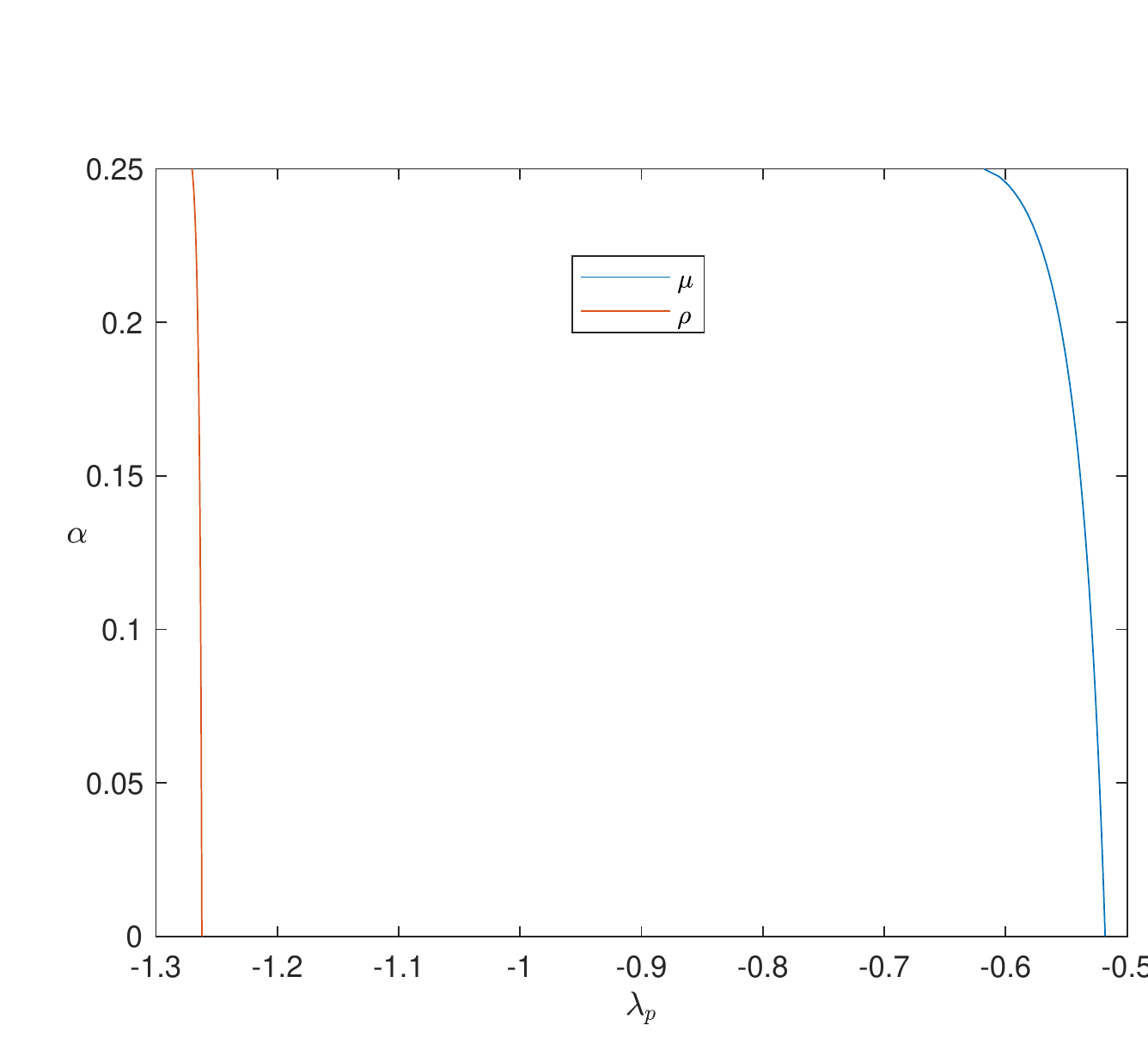}
\caption{
The depedence of the special values of $\lambda_s$ (left panel) and $\lambda_p$ (right panel) on the Gauss-Bonnet coefficient $\alpha$ in the canonical ensemble (red lines) and the grand canonical ensemble (blue lines) in the holographic model from the 4D Einstein-Gauss-Bonnet gravity with $q_s=q_p=1$, $m_s^2=-2/L_{\text{eff}}^2$ and $m_p^2=0$.}\label{alphalambdaSP4D}
\end{figure}

\begin{figure}
\centering
\includegraphics[width=0.49\columnwidth,origin=c,trim=90 260 120 230,clip]{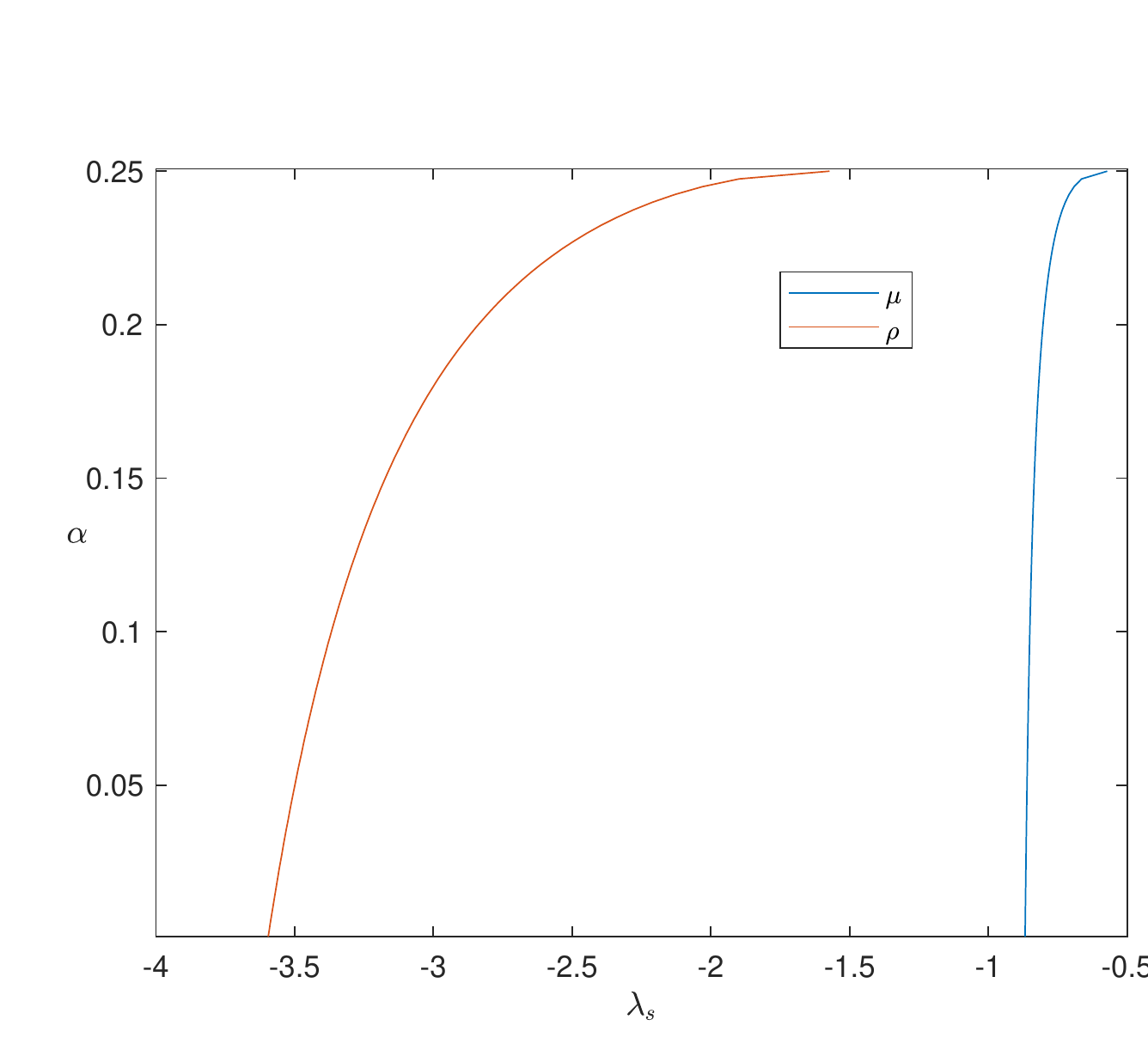}
\includegraphics[width=0.49\columnwidth,origin=c,trim=90 260 120 230,clip]{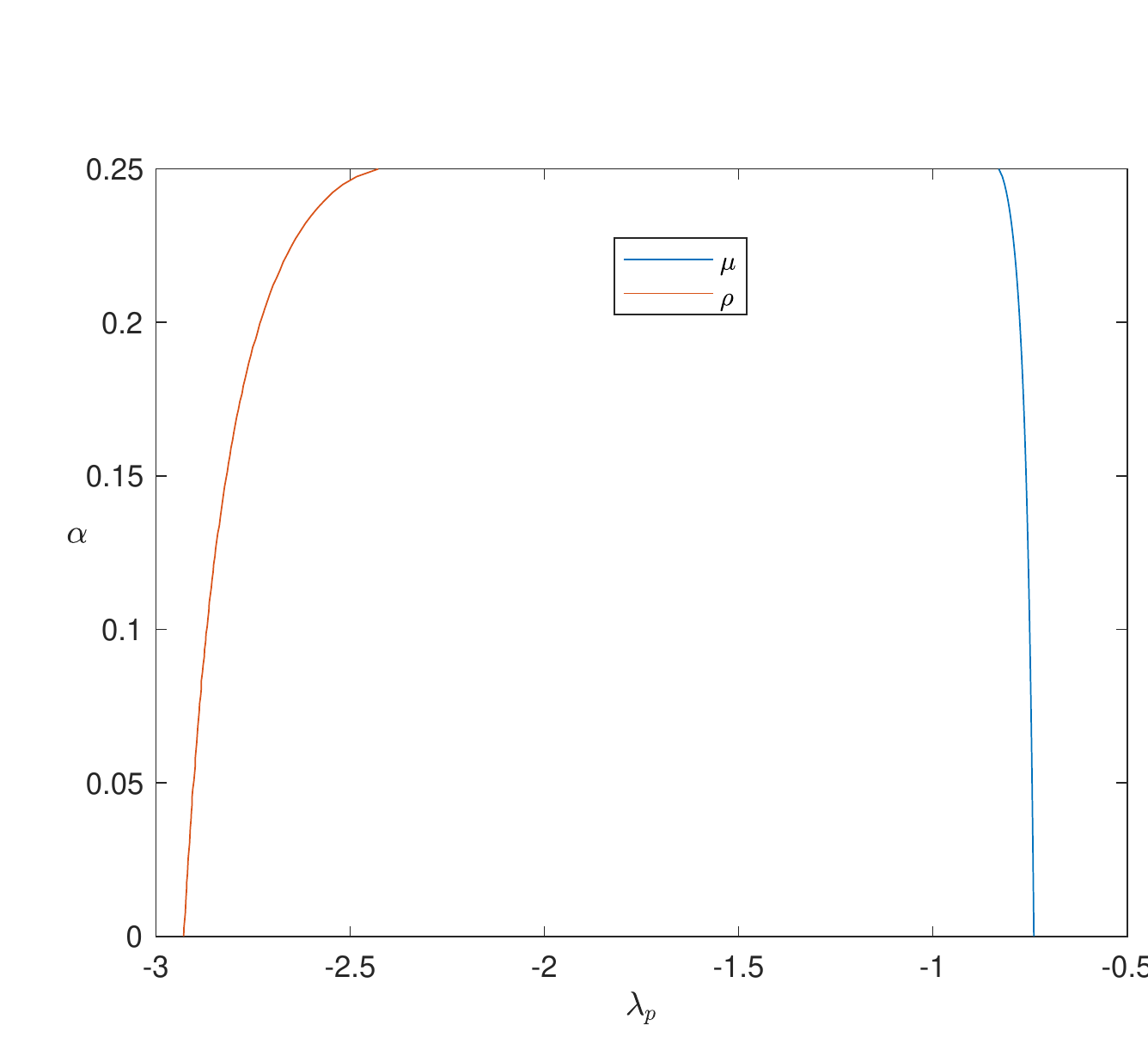}
\caption{
The depedence of the special values of $\lambda_s$ (left panel) and $\lambda_p$ (right panel) on the Gauss-Bonnet coefficient $\alpha$ in the canonical ensemble (red lines) and the grand canonical ensemble (blue lines) in the holographic model from the 5D Einstein-Gauss-Bonnet gravity with $q_s=q_p=1$, $m_s^2=-3/L_{\text{eff}}^2$ and $m_p^2=0$.}\label{alphalambdaSP5D}
\end{figure}

The story for the p-wave order is similar, and we also expect two special values of the coefficient $\lambda_p$. In the Einstein-Gauss-Bonnet gravity, it is natural to ask how these special values of $\lambda_s$ and $\lambda_p$ depend on the Gauss-Bonnet coefficient $\alpha$, therefore we plot the curves in Figs.~\ref{alphalambdaSP4D} and \ref{alphalambdaSP5D} for the cases with 4D and 5D bulks, respectively.

We can see from Fig.~\ref{alphalambdaSP4D} that for both $\lambda_s$ and $\lambda_p$, the special values in the grand canonical ensemble are indeed different and larger than the special values in the canonical ensemble. And the two special values for the s-wave order increase with larger $\alpha$, while the special values of $\lambda_p$ for the p-wave order decrease. From the curves in Fig.~\ref{alphalambdaSP5D} we also see that the special values in the the grand canonical ensemble are larger than the special values in the canonical ensemble. What is different is that in the case with a 5D bulk, the special value of $\lambda_p$ in the canonical ensemble becomes larger along with the increase of $\alpha$. This difference might be a result of different operator dimensions of the different orders in different spacetime dimensions, and would be studied in more detail in the future.

Based on the above curves for special values of $\lambda_s$ and $\lambda_p$, it is more convenient to realize the 1st order phase transitions in the Gauss-Bonnet gravity and study the interesting topics such as the spinodal decomposition.
%\cite{Herzog:2010vz}\cite{Zhang:2021vwp}
%4D
%5D
\section{conclusion and discussion}\label{conclusion}
In this work, we study the effect of the Gauss-Bonnet term with the parameter $\alpha$ on the phase transitions in the holographic s+p model from the 4D and 5D Einstein-Gauss-Bonnet gravity. We first show the $q_p-\mu$ phase diagrams at $\alpha=1\times 10^{-7}$ to choose better values of $q_p$, where the coexistent s+p phase is clearly presented in the phase diagram. Then we plot $\alpha-\mu$ phase diagrams with $q_p=0.9487$ in the case with a 4D bulk and $q_p=0.9798$ in the case with a 5D bulk to show the influence of the higher curvature term on the phase transitions involving the s+p coexistent solution. Based on these two phase diagrams, we increase the values of $q_p$ a little bit to $q_p=0.9695$ in the 4D case and $q_p=0.9950$ in the 5D case, respectively, to show the phase transitions of the coexistent s+p phase near the Chern-Simons limit, where the reentrant s+p phase transition is clearly realizable with appropriate value of $\alpha$ in both the 4D and 5D cases. From these phase diagrams, we also see that the laws of phase transitions in holographic models from the 4D Einstein-Gauss-Bonnet gravity is qualitatively the same with that from the 5D Einstein-Gauss-Bonnet gravity.

In addition, we study the denpendenc of the special values of the coefficients $\lambda_s$ and $\lambda_p$ on the Gauss-Bonnet parameter $\alpha$ in both the canonical and the grand canonical ensembles, below which the single condensate grows leftwards near the critical point. These values are important in realizing the 1st order phase transitions as well as studying the spinodal decomposition in holography.

Based on this study, it is interesting to further explore the full phase structure including the 2nd, 1st and 0th order phase transitions in the 4D and 5D Einstein-Gauss-Bonnet gravity. The Gauss-Bonnet parameter is also possible to be transformed to other physical quantities such as the pressure~\cite{Nie:2015zia} by applying the scaling symmetries. With the influence of the Gauss-Bonnet parameter, we also get more power on tuning the quasi-normal modes (QNMs) as well as the full non-equilibrium dynamics of the system, which is an important field for applying the gauge/gravity duality and needs further investigations in the future.
\section*{Acknowledgements}
ZYN would like to thank Matteo Baggioli, Yu Tian, Hua-Bi Zeng for useful discussions. This work is partially supported by NSFC with Grant No.11965013, 12275079, 12035005, 11881240248, 11565017. ZYN is partially supported by Yunnan High-level Talent Training Support Plan Young $\&$ Elite Talents Project (Grants no. ).

\end{document}